\documentclass[preprint,12pt]{elsarticle}

\usepackage{graphicx}

\usepackage{amssymb}
\usepackage{amsmath}

\usepackage{lineno}

\journal{Nuclear Instruments and Methods in Physics Research B}

\begin{document}

\begin{frontmatter}



\title{Structure of the channeling electrons wave functions under dynamical chaos conditions}


\author[Kharkov1,Kharkov2]{N.F. Shul'ga}
\author[Belgorod]{V.V. Syshchenko\corref{sysh}}
\ead{syshch@yandex.ru}
\author[Belgorod]{A.I. Tarnovsky}
\author[Dubna]{A.Yu. Isupov}

\cortext[sysh]{Corresponding author. Tel.: +7 4722 301819; fax: +7 4722 301012}

\address[Kharkov1]{National Science Center ``Kharkov Institute of Physics and Technology'',\\ 1, Akademicheskaya St., Kharkov 61108, Ukraine}
\address[Kharkov2]{V.N. Karazin National University, 4, Svodody Sq., Kharkov 61022, Ukraine}
\address[Belgorod]{Belgorod National Research University, 85, Pobedy St., Belgorod 308015, Russian Federation}
\address[Dubna]{Laboratory of High Energy Physics, Joint Institute for Nuclear Research,\\ 141980, Dubna, Moscow region, Russian Federation}

\begin{abstract}
The stationary wave functions of fast electrons axially channeling in the silicon crystal near [110] direction have been found numerically for integrable and non-integrable cases, for which the classical motion is regular and chaotic, respectively. The nodal structure of the wave functions in the quasi-classical region, where the energy levels density is high, is agreed with quantum chaos theory predictions.
\end{abstract}


\end{frontmatter}



\section{Introduction}

The motion of fast charged particles through a crystal can be both regular and chaotic  from the classical viewpoint (see, e.g., \cite{AhSh, AhSh2}). In quantum mechanics the chaoticity of classically non-integrable Hamiltonian systems manifests itself in different ways, both in the statistical properties of the energy spectra and the structure of stationary wave functions of individual quantum states \cite{Gutz, Schuster}. For the latter case the chaoticity is especially evident for two-dimensional systems where the patterns formed by the nodal lines $\psi (x,y) = 0$ of the Hamiltonian eigenfunctions are quite different in integrable and non-integrable (regular and chaotic) cases.

\section{Axial channeling from the classical mechanics viewpoint}

\begin{figure}
\begin{center}
\includegraphics[width=\columnwidth]{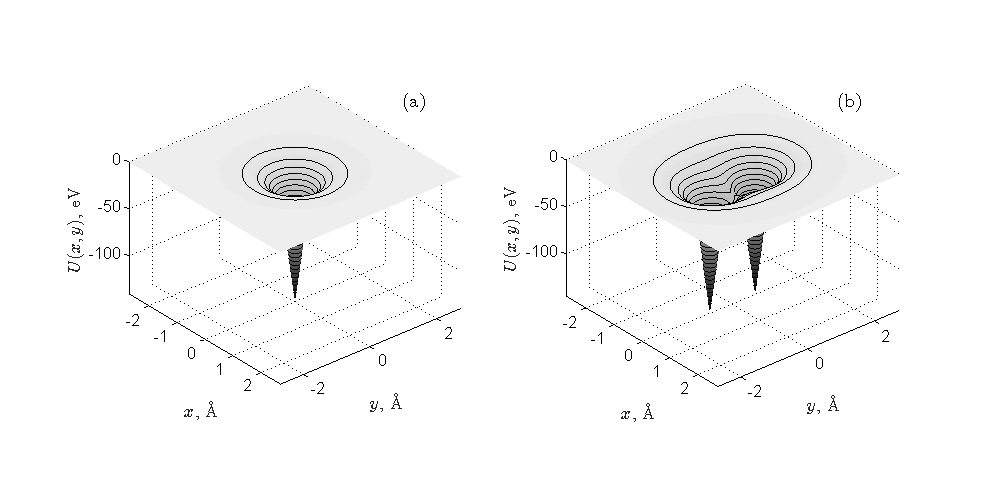}
\end{center}
\caption{Electron potential energy in the continuous potentials of the stand-alone atomic string (a) and two neighbor [110] atomic strings (b) of the silicon crystal. The potential parameters are found in \cite{AhSh}.}\label{Fig1}
\end{figure}

\begin{figure}
\begin{center}
\includegraphics[width=\columnwidth]{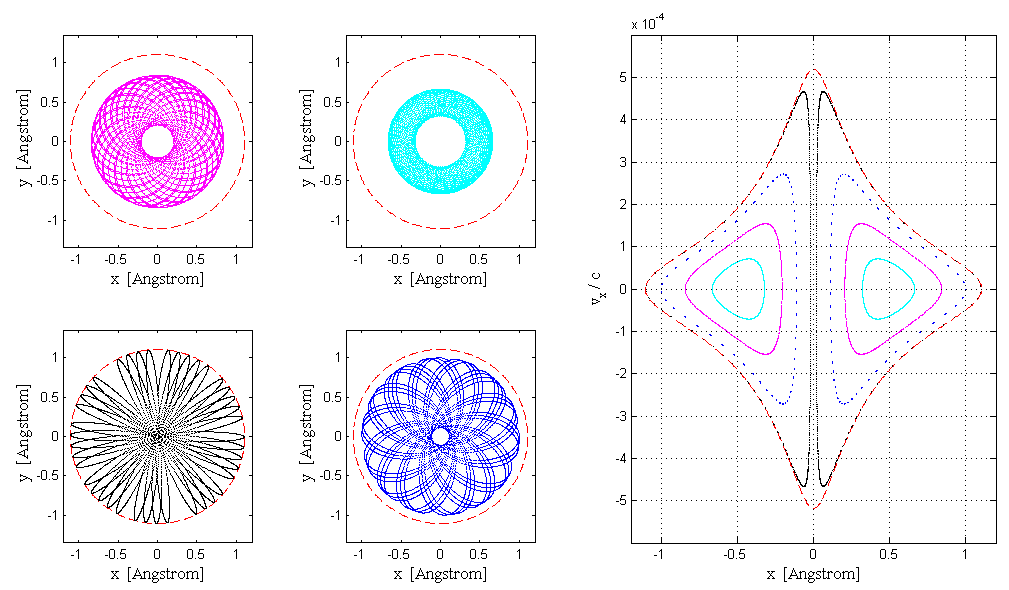}
\end{center}
\caption{{\it(Color online).} Sample $E_\parallel = 1$ GeV, $E_\perp = -6.064$ eV electron's trajectories in the continuous potential of the stand-alone [110] atomic string of the silicon crystal (left) and the corresponding Poincar\'e sections for $y^* = 0$ (right). Dashed lines (red online) correspond to the classical border of motion $U(x,y) = E_\perp$ (on the trajectories plots) and $U(x,y^*) + E_\parallel v_x^2/2c^2 = E_\perp$ (on the Poincar\'e plot).}\label{Fig2}
\end{figure}

\begin{figure}
\begin{center}
\includegraphics[width=\columnwidth]{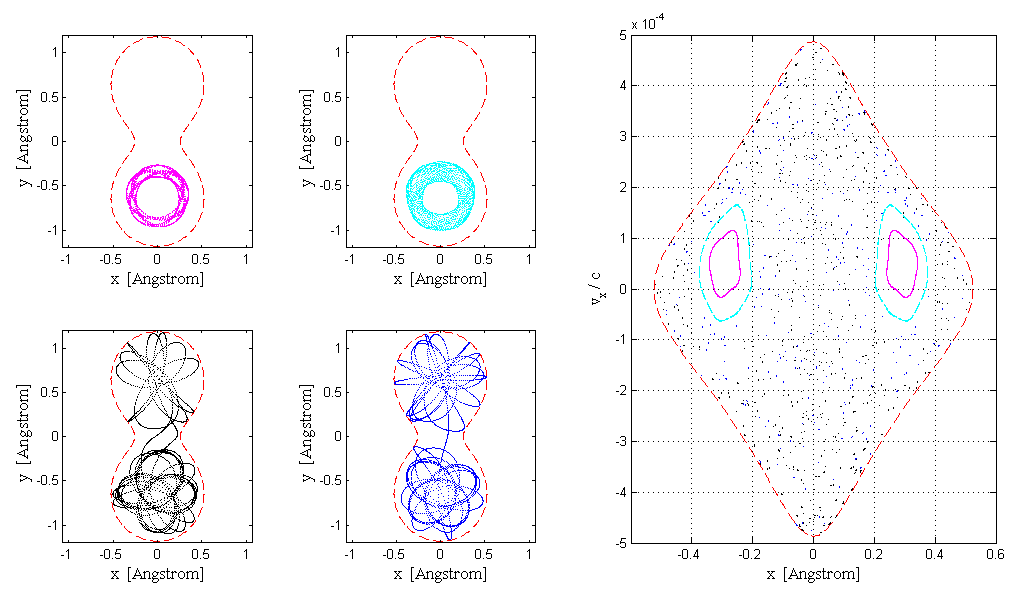}
\end{center}
\caption{The same as in Fig. \ref{Fig2} for the double-well potential of two neighbor atomic strings, $E_\perp = -26.151$ eV and $y^* = -a/8$, where $a=5.431$ \AA \ is the silicon lattice period.}\label{Fig3}
\end{figure}

\begin{figure}
\begin{center}
\includegraphics[width=\columnwidth]{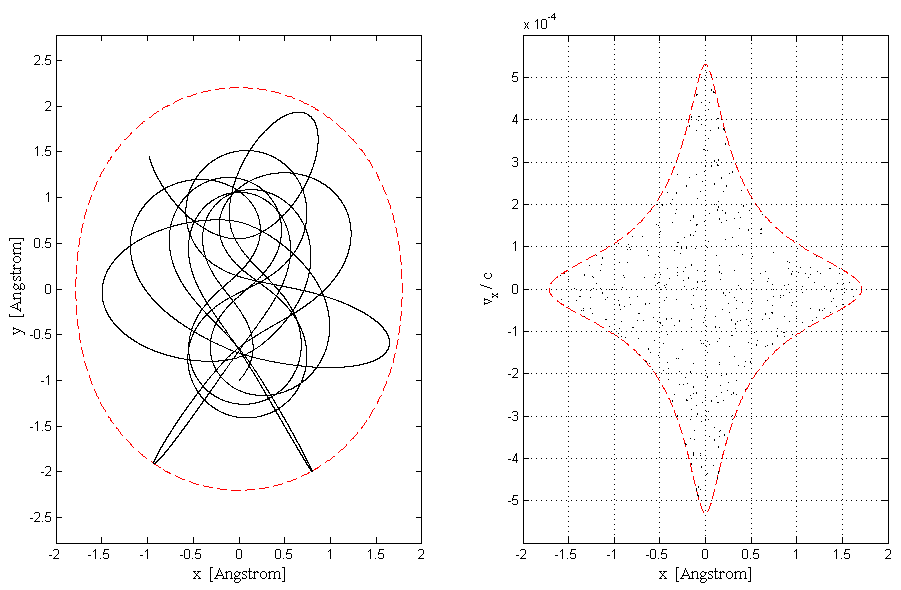}
\end{center}
\caption{
Same as in Fig.~\ref{Fig3} for $E_{\perp} = -4.246$~eV.}\label{Fig4}
\end{figure}

The problem of the particles' motion through a crystal could be reduced to two-dimensional when the particles are incident under small angle to one of the crystallographic axes
densely packed with atoms (see, e.g., \cite{AhSh, AhSh2, Ugg}). The potential of the
atomic string in this case can be effectively replaced by continuous potential
averaged along the string axis $z$. The longitudinal
component $p_\parallel = p_z$ of the particle's momentum is conserved due to
continuous potential independence on $z$.
The transversal finite motion in such potential (so called axial channeling)
is described by the two-dimensional equation of motion
\begin{equation}\label{ssti.1}
    \frac{E_\parallel}{c^2} \frac{d^2\mathbf r}{dt^2} = - \nabla U(x,y),
\end{equation}
where $\mathbf r = (x,y)$ is the two-dimensional radius-vector of the channeling electron in the transverse plane, and the parameter $E_\parallel /c^2$ (where $E_\parallel = (m^2c^4 + p_\parallel^2c^2)^{1/2}$ is the longitudinal motion energy) plays the role of the particle mass \cite{AhSh}.

The two-dimensional motion described by Eq. (\ref{ssti.1}) always possesses
one integral of motion --- the transverse motion energy
\begin{equation}\label{ssti.2}
    E_\perp = \frac{E_\parallel}{2c^2} (v_x^2 + v_y^2) + U(x,y),
\end{equation}
where $v_x = dx/dt$ and $v_y = dy/dt$ are the electron's velocity components in the $(x,y)$ plane.

It is well known from the classical mechanics (see, e.g., \cite{LL1}) that the axial symmetry of the potential, $U(x,y) = U(r)$ ($r = \sqrt{x^2+y^2}$ is the length
of the electron's radius-vector) leads to the conservation of the angular momentum projection on the potential axis
\begin{equation}\label{ssti.3}
    M = \frac{E_\parallel}{c^2} r^2 \frac{d\varphi}{dt},
\end{equation}
where $\varphi = \arctan (y/x)$ is the electron's polar angle in the $(x,y)$ plane. The presence of two integrals of motion, (\ref{ssti.2}) and (\ref{ssti.3}), gives the possibility to separate variables in the equation of motion (\ref{ssti.1}) and integrate it in quadratures. Note that the radial motion is reduced in this case to one-dimensional problem of motion in the effective potential
\begin{equation}\label{ssti.4}
    U_{\mbox{\footnotesize eff}}(r) = U(r) + \frac{M^2c^2}{2E_\parallel r^2},
\end{equation}
where the second term is called the centrifugal energy.

The presence or absence of the second integral of motion (in addition to the transverse energy (\ref{ssti.2}) which is always present) under some conditions can be tested numerically using Poincar\'e section method \cite{AhSh, Gutz, Schuster}. For the given value of one coordinate, say $y=y^*$, the Cartesian coordinates and components of the electron's velocity are related through the equation
\begin{equation}\label{ssti.5}
    E_\perp = E_\perp (x, y^*, v_x, v_y).
\end{equation}
The presence of some second integral of motion $J$ means the existence of the another relation
\begin{equation}\label{ssti.6}
    J_\perp = J_\perp (x, y^*, v_x, v_y).
\end{equation}
The equations (\ref{ssti.5}) and (\ref{ssti.6}) define implicitly the function
\begin{equation}\label{ssti.7}
    v_x = v_x(x; E_\perp , J, y^*)
\end{equation}
of the single argument $x$ and three parameters $E_\perp$, $J$, and $y^*$. So, if the points that correspond $y=y^*$ on the plane $(x,v_x)$ for the electron's trajectory with the given initial values form the smooth curve, that reflects the existence of the function (\ref{ssti.7}) and hence the integral of motion (\ref{ssti.6}) and the integrability of the equation of motion (\ref{ssti.1}). In the opposite case the points on the Poincar\'e plot looks like randomly spread.

The equation of motion non-integrability is closely related to the chaoticity of motion, when the deterministic trajectories become unstable and random-looking.

The continuous potential of the stand-alone atomic string (Fig. \ref{Fig1} (a)) is just the case when the presence of two conserving values --- the transverse motion energy $E_\perp$ and the particle's angular momentum projection on the axis of symmetry --- leads to integrability of two-dimensional mechanical system and regularity of its motion. The sample trajectories of the electron in that potential as well as corresponding Poincar\'e plots are presented in Fig. \ref{Fig2}.

The presence of the neighbor strings in the crystal destroys the axial symmetry of the potential and breaks the angular momentum conservation, so the equation of motion appears non-integrable. The simplest example is the diamond-like crystals where [110] strings are grouped into pairs, and the influence of far-away strings on the particle's motion could be neglected in some approximation (Fig. \ref{Fig1} (b)). The classical dynamics of the particle in this field is chaotic for the most part of the initial values above the potential's saddle point \cite{AhSh, AhSh2}. The sample trajectories of the electron in that potential as well as corresponding Poincar\'e plots are presented in Figs. \ref{Fig3} (near the saddle point) and \ref{Fig4} (in the potential upper part).

\section{Wave functions in the quantum theory of axial channeling}

The transverse motion of channeling electrons can be quantized \cite{AhSh}. The electron's stationary states are described by the two-dimensional Schr\"odinger equation
\begin{equation}\label{ssti.8}
    \hat H \psi(x,y) = E_\perp \psi(x,y),
\end{equation}
where the Hamiltonian can be written in the form
\begin{equation}\label{ssti.9}
    \hat{H} = - \frac{\hbar^2}{2 E_\parallel / c^2} \left[ \frac{1}{r} \frac{\partial}{\partial r} \left( r \, \frac{\partial}{\partial r} \right) + \frac{1}{r^2} \frac{\partial^2}{\partial \varphi^2} \right] + U(r)
\end{equation}
for the stand-alone atomic string with the axially symmetric potential. This symmetry permits to separate variables in Eq. (\ref{ssti.8}) and integrate the angular part that leads to the following form of the Hamiltonian (\ref{ssti.9})  eigenfunctions:
\begin{equation}\label{ssti.10}
    \psi_{n_r, m} (r, \varphi) = \frac{1}{\sqrt{2 \pi}} \, e^{i m \varphi} \rho_{n_r, |m|} (r) .
\end{equation}
Here the $\rho_{n_r, |m|} (r)$ function is the solution of
\begin{equation}\label{ssti.11}
    \frac{1}{r} \frac{\partial}{\partial r} \left( r \, \frac{\partial}{\partial r} \right) \rho_{n_r, |m|} (r) +
\end{equation}
\begin{equation*}
    + \frac{2 E_\parallel}{\hbar^2c^2} \left( E_\perp - U_{\mbox{\footnotesize eff}}(r) \right) \rho_{n_r, |m|} (r) = 0,
\end{equation*}
where the effective potential energy is given by Eq. (\ref{ssti.4}) with $M = \hbar m$. We see that the electron quantum states in the axially-symmetric potential $U(r)$ are classified by two quantum numbers: the radial one $n_r$ and projection $m$ of the orbital momentum to the symmetry axis of the field. The $n_r$ is equal to the number of zeroes of the function $\rho_{n_r, |m|} (r)$ at finite $r$ (except the possible zero at $r = 0$). The states with $m = 0$ are non-degenerated, while ones with non-zero $m$ are doubly degenerated by the signum of $m$.

The numerical method we used to find the channeling electron's stationary wave functions is the so called spectral method \cite{Feit82}. It is equally good for the potentials with and without axial (or any other) symmetry. This method had been applied to the channeling problem for the first time in \cite{Dabagov1}--\cite{Dabagov88}. Some technical details of our computations are outlined in the \ref{AppA}.

\begin{figure}
\begin{center}
\includegraphics[scale=0.65]{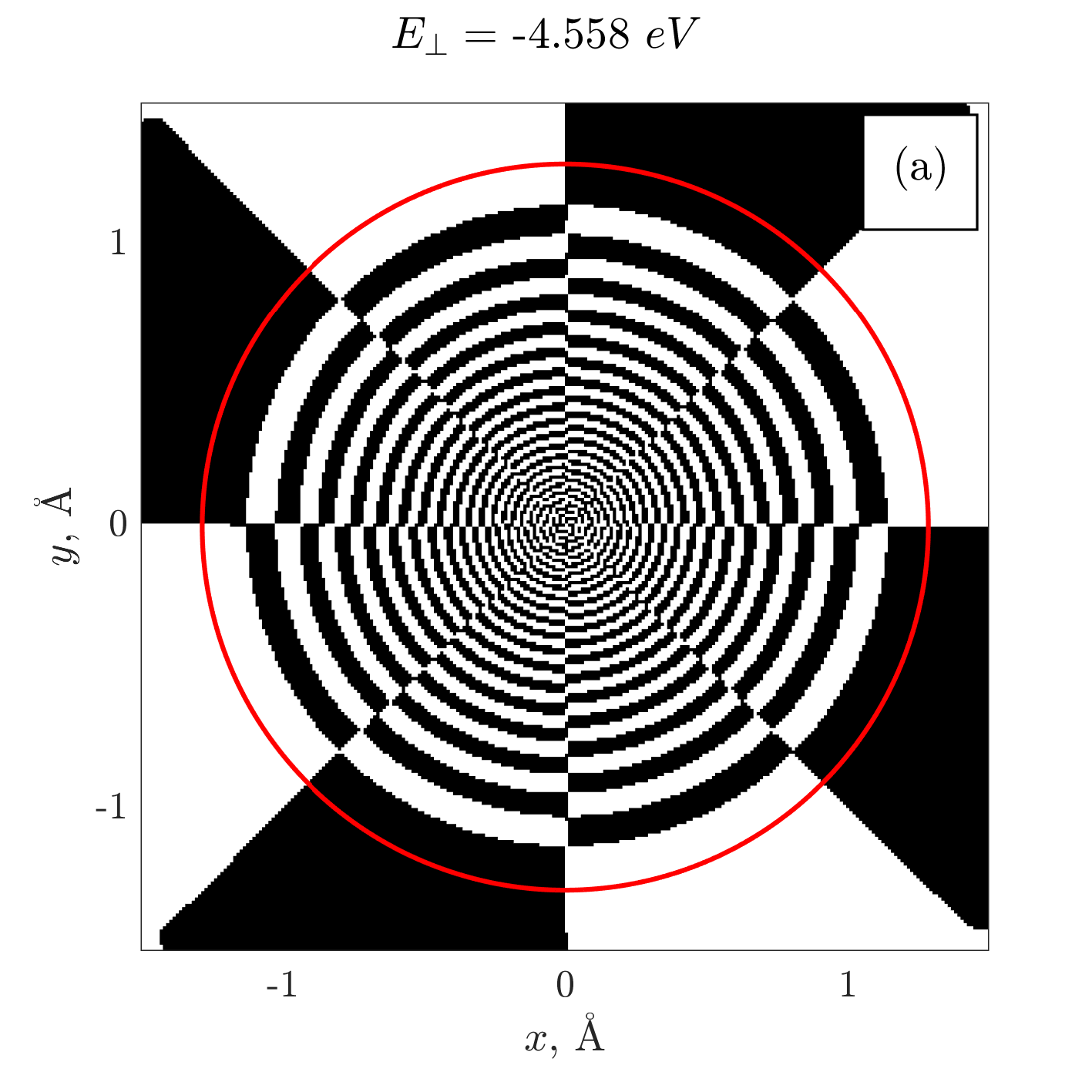} \\
\includegraphics[scale=0.65]{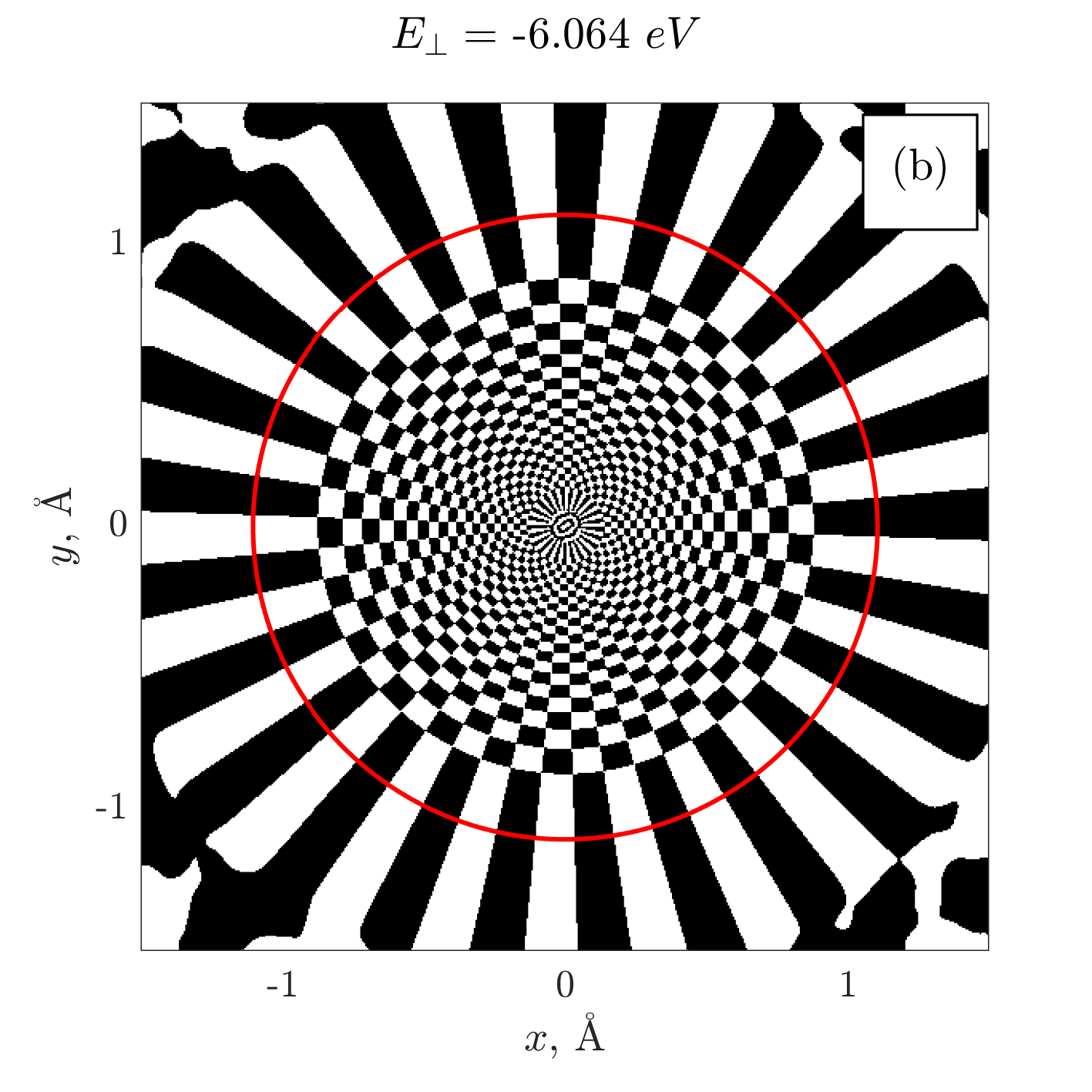}
\end{center}
\caption{{\it(Color online).} Patterns formed by positive (white) and negative (black) domains of the Hamiltonian eigenfunction for $E_\parallel = 1$ GeV electron in the single-string potential well for two different transverse energy $E_\perp$ eigenvalues. Dashed lines (red online) are the classical turning lines $U(x,y) = E_\perp$. The irregularity on the periphery is due to numerical uncertainties in the eigenfunction computation.}\label{Fig5}
\end{figure}

\begin{figure}
\begin{center}
\includegraphics[width=\columnwidth]{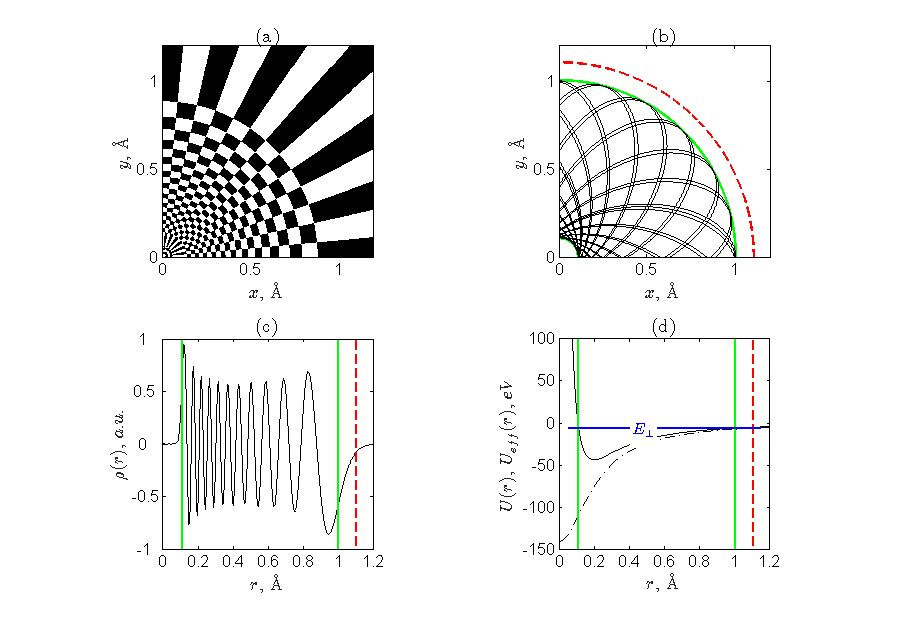} \\
\end{center}
\caption{{\it(Color online).} The correspondence between quantum (a) and classical (b) views on the channeling electron's motion in the stand-alone string continuous potential under the conditions of Fig. \ref{Fig5} (b). The radial part of the wave function $\rho_{n_r, |m|} (r)$  (c) is extracted from the two-dimensional wave function $\psi(x,y)$ by tracing it out along the radius. The potential energy $U(r)$ (dash-dot curve) and the effective potential energy (\ref{ssti.4}) (solid curve) are plotted in the panel (d). The classical borders of motion $U_{\mbox{\footnotesize eff}}(r) = E_\perp$ are presented as solid (green online) lines; dashed lines (red online) correspond to $U(r) = E_\perp$.}\label{Fig6}
\end{figure}

The spectral method could not discriminate the eigenstates degenerated by the signum of $m$, and returns the solutions of the Schr\"odinger equation in the form
\begin{equation}\label{ssti.12}
    \psi_{n_r, m} (r, \varphi) \propto \rho_{n_r, |m|} (r) \cos \left[ \frac{}{}\! |m| \varphi + \alpha_m \right]
\end{equation}
($\alpha_m$ is some constant phase) which is the superposition of the functions (\ref{ssti.10}) with positive and negative $m$. Note that the eigenfunctions of the form (\ref{ssti.12}) are real, not complex. This fact reflects the existence of general theorem which says the complete set of basis functions of a real Hamiltonian can be taken as real without loss of generality (see, e.g., \cite[sec. 3.1.2]{Stockmann}). The convenient way of graphical presentation of such real wave functions is to paint the domains $\psi (x,y) < 0$ and $\psi (x,y) > 0$ on the $(x,y)$ plane in black and white, respectively. The sample eigenfunctions of the electron channeling in the axial symmetric potential are presented in Figs. \ref{Fig5}, \ref{Fig6}.

The nodal lines of the wave function $\psi (x,y) = 0$ cross each other due to the variables separability in the Schr\"odinger equation, and form the well-recognizable checkerboard-like pattern. The radial $n_r$ and orbital $m$ quantum numbers that characterize the given state could be found easily by counting the nodal lines of the corresponding wave function. For example, the wave function in Fig. \ref{Fig6} corresponds to $m=25$; this value is used to calculate the centrifugal energy and plot the curve of the effective potential energy (\ref{ssti.4}) in Fig. \ref{Fig6} (d).

In the classically prohibited regions $U_{\mbox{\footnotesize eff}}>E_\perp$ the radial part of the wave function $\rho_{n_r, |m|} (r)$ has no zeros, so there are no circular nodal lines in these regions, only the straight radial ones. Note that the centrifugal barrier near the string axis also forms the classically prohibited domain, well recognizable by switching from the checkerboard-like pattern to the straight nodal lines. The \cite{Berezovoj04} states this switching happens just on the classical turning line $U(x,y) = E_\perp$. However, in fact it happens on some distance before it, and just before the line $U_{\mbox{\footnotesize eff}} = E_\perp$, as could be seen in Fig. \ref{Fig6}. Indeed, the solution of the radial part of the Schr\"odinger equation (\ref{ssti.11}) would be the function that monotonically tends to zero in the classically forbidden regions (under $r\to \infty$ and $r\to 0$), similarly to the problem of the particle in one-dimensional potential well. Such function could not be zero in the turning point.

The angular momentum conservation is violated in double-well potential that leads to the motion equation non-integrability and the wave functions structure dramatic change. The qualitative distinctions between the wave functions in the regular and chaotic cases have been discovered and studied by various authors (see, e.g. \cite{Gutz} and references therein, and, especially, \cite{Bies, Stratt, Berezovoj04}). They can be summarized (in application to our problem of channeling) in two groups:
\begin{itemize}
  \item[(i)] the nodal lines of the regular wave function exhibit crossings (in separable case) or very tiny quasi-crossings (in non-separable, but still regular case, see \cite{Stratt}) forming checkerboard-like pattern; the nodal lines of the chaotic wave function form a sophisticated pattern of black and white islands, the nodal lines quasi-crossings have significantly larger avoidance ranges;
  \item[(ii)] near the classical turning line the nodal structure of the regular wave function immediately switches to the straight nodal lines, in the outer domain going to infinity; for the chaotic wave function an intermediate region exists outside the turning line, where some of the nodal lines pinch-off, making transition to the classically forbidden region more graduate and not so manifesting in the nodal structure.
\end{itemize}
Note that the quantum chaos manifestations have to be studied in the quasi-classical domain, where the energy levels density is high enough to be close to the classical situation of continuous energy spectrum. The total number of energy levels grows with the $E_\parallel$ increase, as it is predicted from semiclassical arguments \cite{AhSh}, that is the reason of the choice $E_\parallel = 1$ GeV for our studies. On the other hand, the maximal level density is reached in the upper part of our expanding potential wells (Fig. \ref{Fig1}).
Hence we had chosen the transverse motion energy domain near the top of the potential well.

The considered two-well potential (Fig. \ref{Fig1}, (b)) has two planes of mirror symmetry: $x = 0$ and $y = 0$. Hence, every eigenstate of the electron in this potential belongs to one of four symmetry classes:
\begin{equation}\label{ssti.13}
\left\{
\begin{aligned}
\psi_{++} (-x,y) = \psi_{++} (x,y), \\
 \psi_{++} (x,-y) = \psi_{++} (x,y), \\
\end{aligned}
\right.
\end{equation}
\begin{equation}\label{ssti.14}
\left\{
\begin{aligned}
\psi_{+-} (-x,y) & = \psi_{+-} (x,y), \\
\psi_{+-} (x,-y) & = -\psi_{+-} (x,y), \\
\end{aligned}
\right.
\end{equation}
\begin{equation}\label{ssti.15}
\left\{
\begin{aligned}
\psi_{-+} (-x,y) & = -\psi_{-+} (x,y), \\
\psi_{-+} (x,-y) & = \psi_{-+} (x,y), \\
\end{aligned}
\right.
\end{equation}
\begin{equation}\label{ssti.16}
\left\{
\begin{aligned}
\psi_{--} (-x,y) = -\psi_{--} (x,y), \\
\psi_{--} (x,-y) = -\psi_{--} (x,y). \\
\end{aligned}
\right.
\end{equation}

\begin{figure}
\begin{center}
\includegraphics[scale=0.5]{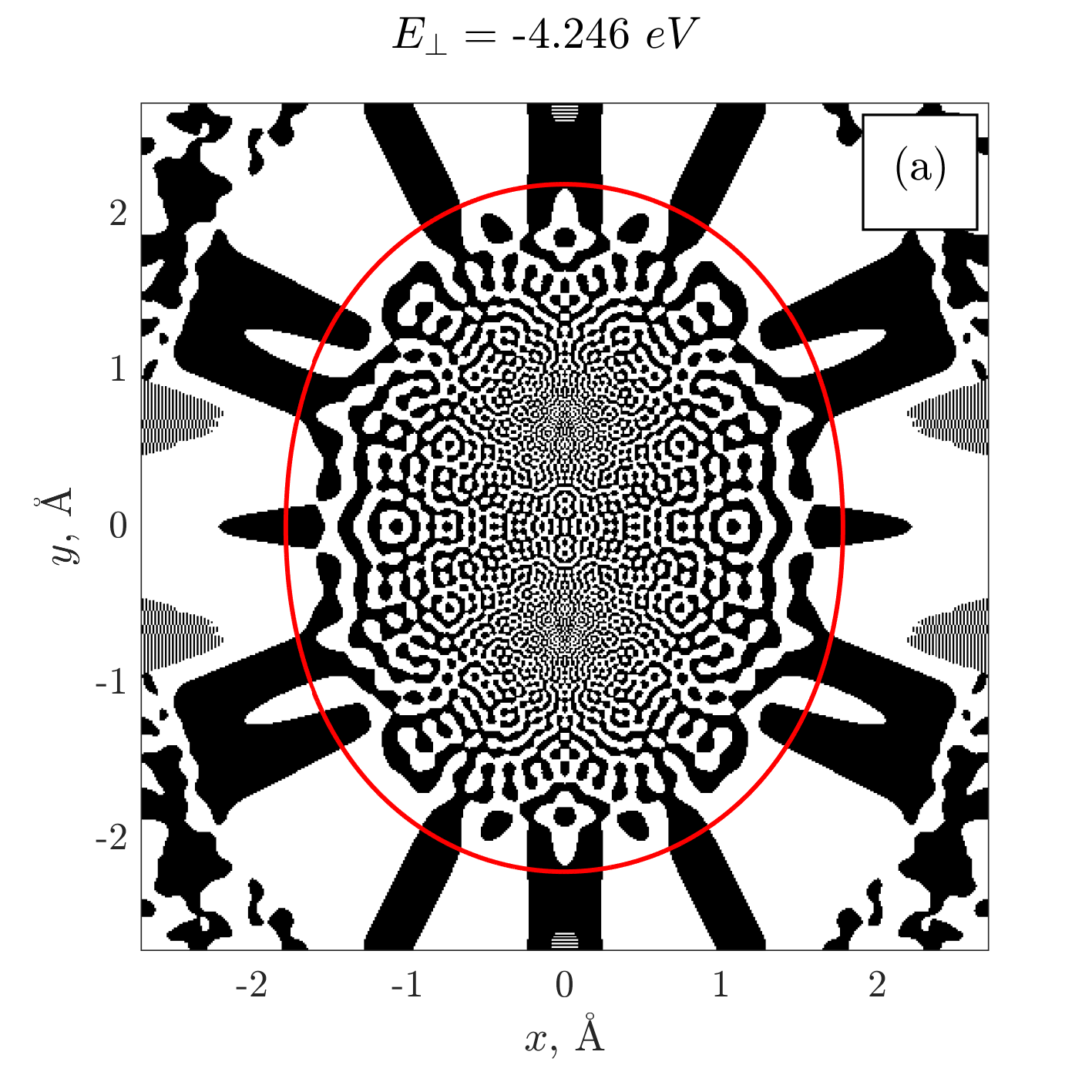} \ \includegraphics[scale=0.5]{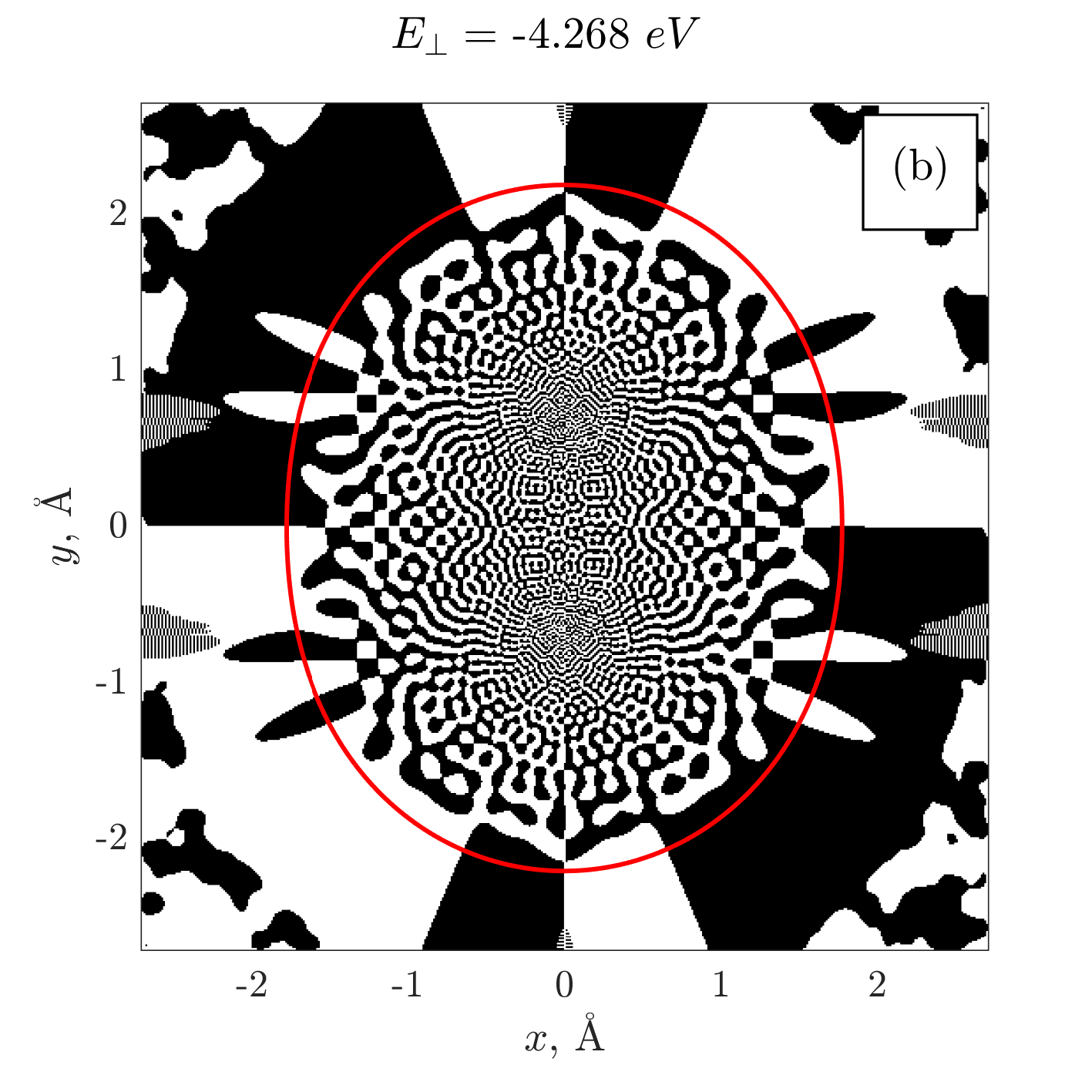} \\
\includegraphics[scale=0.5]{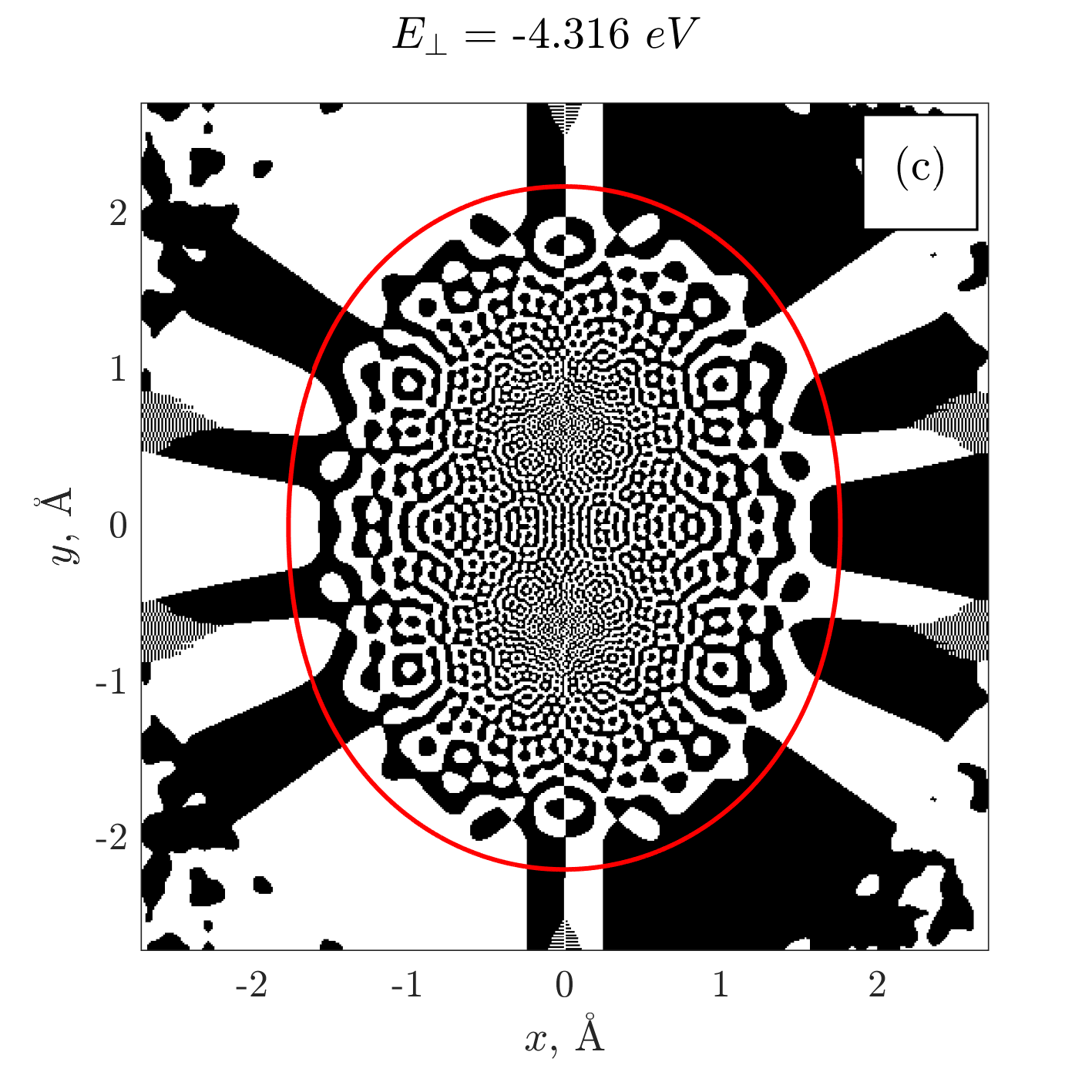} \ \includegraphics[scale=0.5]{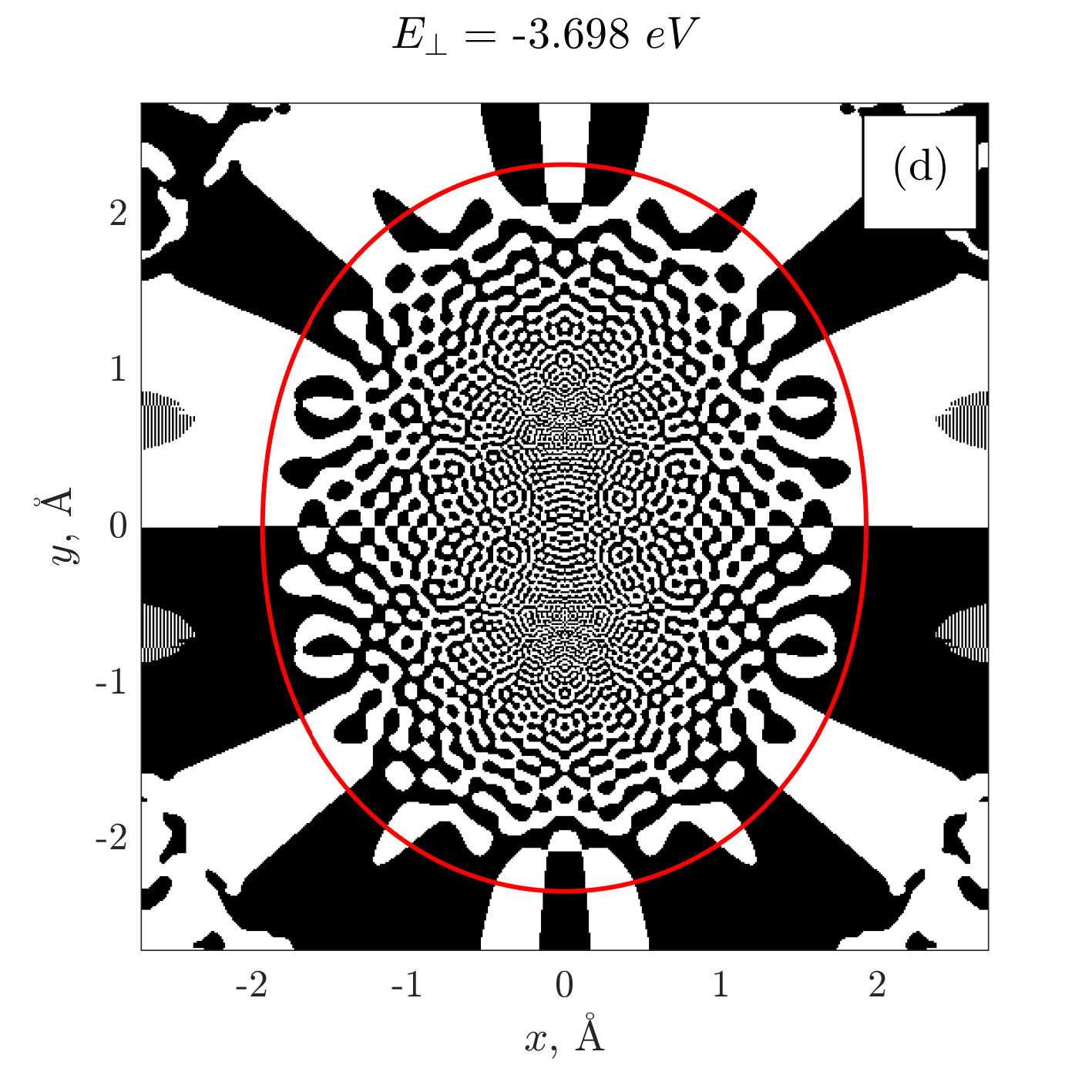} \\
\end{center}
\caption{The same as in Fig. \ref{Fig5} for the case of the double well.}\label{Fig7}
\end{figure}

Some of the wave functions in double-well potential are plotted in Fig. \ref{Fig7}. We see a beautiful correspondence to the predictions (i) and (ii) of the quantum chaos theory for the eigenfunctions of every of four symmetry classes (\ref{ssti.13})--(\ref{ssti.16}).

Each of our patterns possesses a large variety of size and density of white and black spots, in contrast to the cases considered by the previous authors. This fact reflects the feature of the potential under consideration, which rapidly deepens near the atomic strings axes. The increase of the frequency of the black and white domains alternation near the axes is due to the local decrease of de Broglie wavelength of the channeling electron's transverse motion in that regions. However, we see that the avoidance of the nodal lines quasi-crossing typical to the quantum chaos persists for the whole pattern and does not depend on the nodal lines local density.

\begin{figure}
\begin{center}
\includegraphics[scale=0.5]{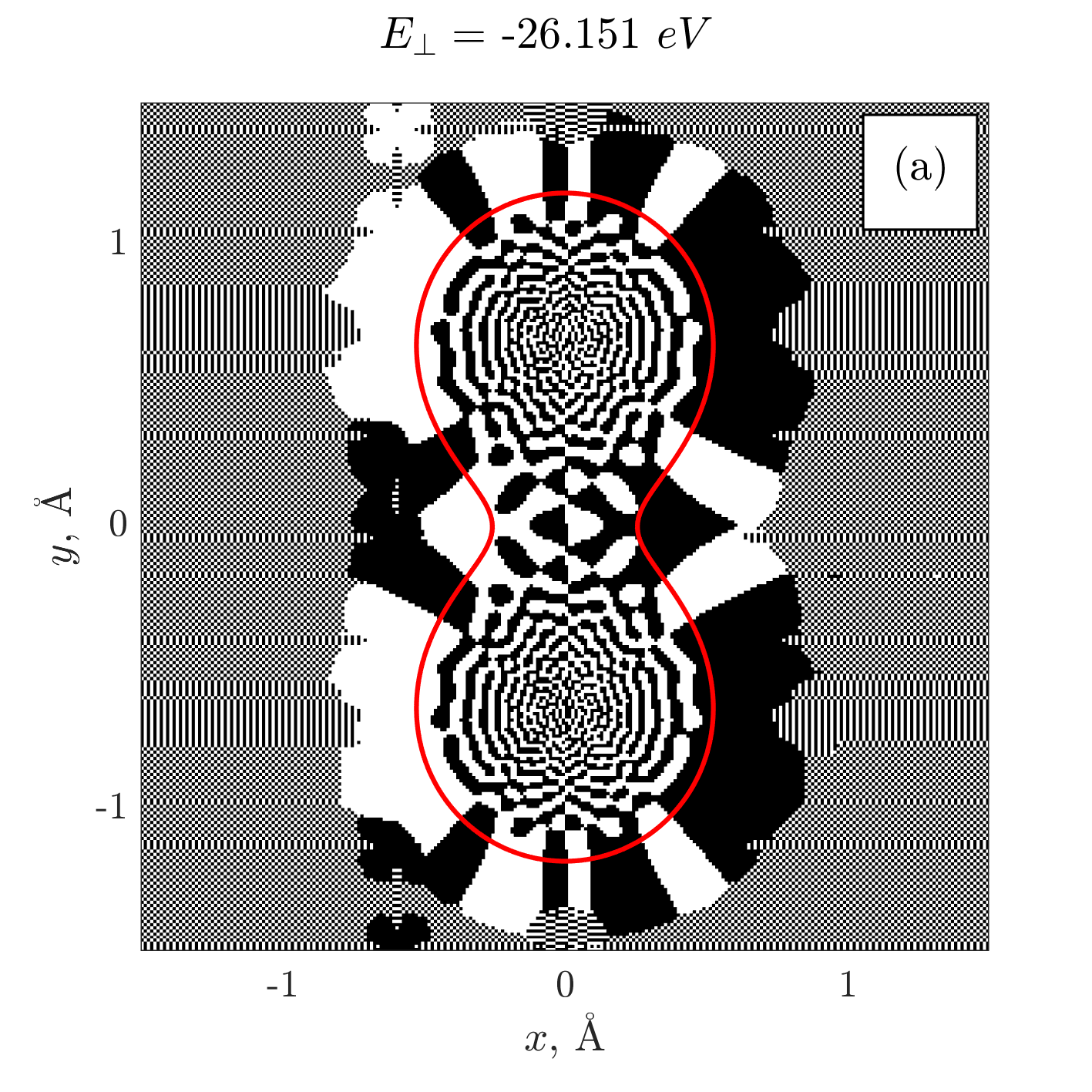} \ \includegraphics[scale=0.5]{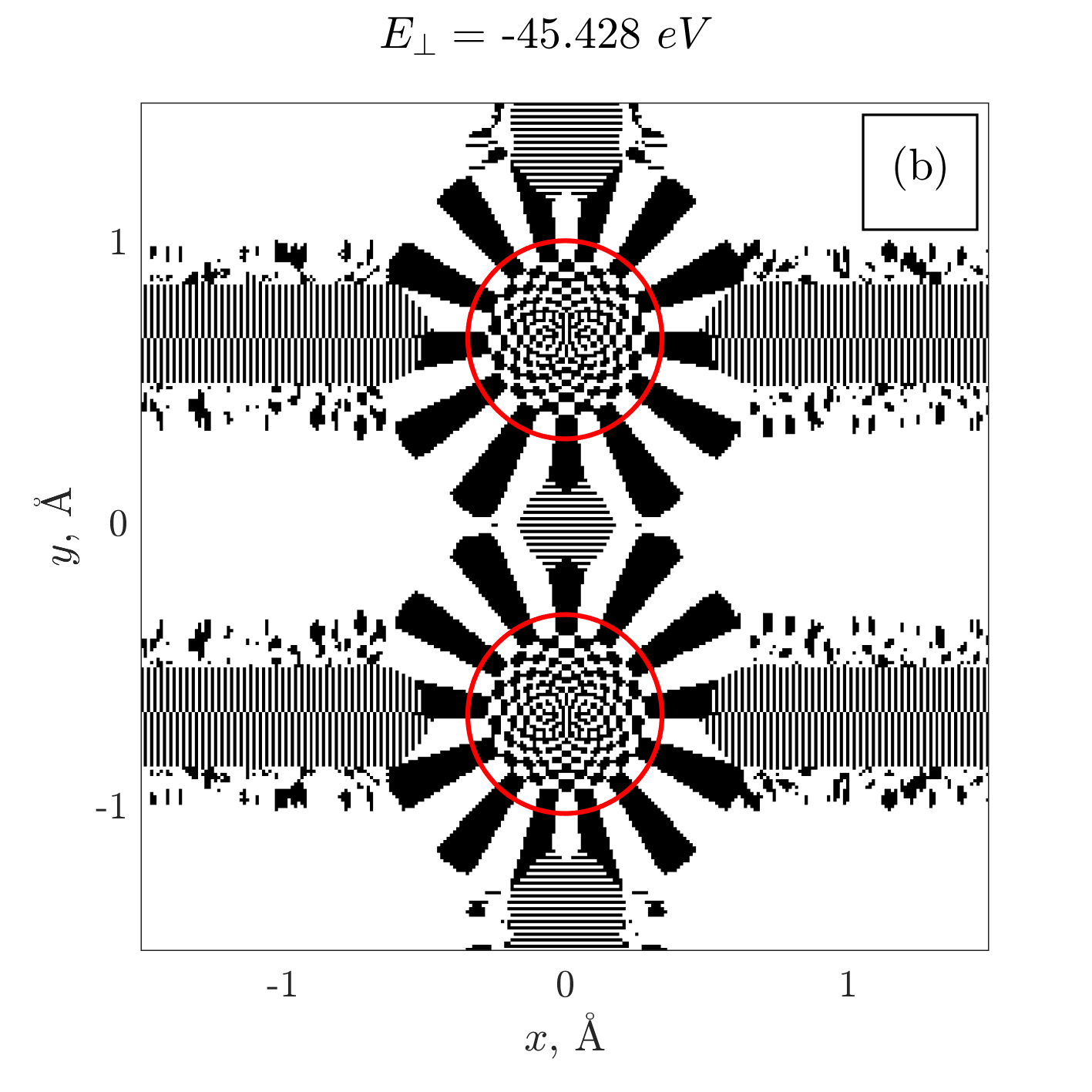} \\
\end{center}
\caption{The same as in Fig. \ref{Fig7} for the $E_\perp$ eigenvalues near the saddle point (a) and deep in the well (b).}\label{Fig8}
\end{figure}

If we deepen into the well we leave the semiclassical domain, and the features of chaotic wave functions outlined above become less manifesting (Fig. \ref{Fig8} (a)). Sinking deeper into the well, we find the axial symmetry of each of the two single potentials restored. That leads to the regularity of motion with corresponding structure of the wave function (Fig. \ref{Fig8} (b)).

\section{Conclusion}

The axial channeling of the high energy electrons ($E_\parallel = 1$ GeV) in the continuous potential of [110] atomic strings of silicon crystal has been considered from both classical and quantum mechanics viewpoints.

The uniform potential of the stand-alone atomic string possesses axial symmetry that leads to the integrability of the equation of motion and regularity of the classical trajectories. The axial symmetry vanishes in the potential of two neighbor atomic strings; the motion in this case is chaotic for the major part of the initial conditions above the potential's saddle point. The chaotic character of the classical motion had been demonstrated using Poincar\'e sections method.

The wave functions that describe the transverse motion eigenstates of the channeling electrons in the quantum approach had been computed using the spectral method based algorithm. We have demonstrated that the nodal structure of the eigenfunctions, which correspond to chaotic motion in classical mechanics, is different qualitatively from one in the regular motion case in the agreement with the quantum chaos theory predictions. These predictions are

--- in the domain classically allowed for motion: nodal lines crossings in the regular case vs non-crossing lines, some of them closed, in the chaotic case;

--- in the classically prohibited domain: direct nodal lines in the regular case vs some nodal lines pinch-off in the chaotic case.

Note that these features had been studied earlier mainly for few simple model systems, like billiards with hard walls and coupled non-linear oscillators. The similar researches for the motion in natural potentials with smooth walls are very rare (see \cite{Berezovoj04} and references therein). We found (in contrast to billiards with plane bottom) substantial size variation of the spots formed by the wave function nodal lines that reflects the variation of the potential well depth across the classically allowed domain.

The switching of the nodal structure from checkerboard-like to straight radially-directed near the string axis in the regular case indicates clearly the centrifugal barrier presence.

Note also that the quantum chaos manifestations in the statistical properties of the energy spectra of the axially channeling electrons have been studied in \cite{Shulga.a, Shulga.b} for $E_\parallel = 400$ MeV and 500 MeV, respectively. The corresponding results for the case $E_\parallel = 1$ GeV under consideration are presented in \ref{AppB}.

\section*{Acknowledgements}

This research is supported in part by the grant of Russian Science Foundation (project 15-12-10019). For one of us (N.F. Shul'ga) the work is partially supported by the Ministry of Education and Science of Ukraine, project No 1-13-15, and National Academy of Science of Ukraine,  project No F12-2015.

\appendix

\section{Overview of numerical method}\label{AppA}

In the spectral method \cite{Feit82} the numerical evolution of the wave packet $ \Psi (x, y, t)$ according to the time-dependent Schr\"odinger equation
\begin{equation}\label{ssti.17}
    \hat{H} \Psi (x, y, t) = i \hbar \, \frac{\partial}{\partial t} \Psi (x, y, t)
\end{equation}
is simulated to find the Hamiltonian eigenvalues and eigenfunctions. Indeed, every solution of the time-dependent Schr\"odinger equation could be expressed as the superposition
\begin{equation}\label{ssti.18}
\Psi (x,y,t) = \sum_{n,j} A_{n,j} \psi_{n,j}(x,y) \exp(-iE_n t/\hbar)
\end{equation}
of the Hamiltonian eigenfunctions $\psi_{n,j}(x,y)$, where the $j$ index is used to distinguish the degenerate eigenstates corresponding to the energy $E_n$. Computation of the correlation function
\begin{equation}\label{ssti.19}
P(t) = \int_{-\infty}^\infty \int_{-\infty}^\infty \Psi^* (x,y,0) \Psi (x,y,t) \, dxdy
\end{equation}
for superposition (\ref{ssti.18}) gives
\begin{equation*}
P(t) = \sum_{n,n',j,j'} \exp(-iE_{n'} t/\hbar) A^*_{n,j} A_{n',j'} \times
\end{equation*}
\begin{equation*}
\times \int_{-\infty}^\infty \int_{-\infty}^\infty \psi^*_{n,j}(x,y) \psi_{n',j'}(x,y) dxdy =
\end{equation*}
\begin{equation*}
    = \sum_{n,n',j,j'} \exp(-iE_{n'} t/\hbar) A^*_{n,j} A_{n',j'} \delta_{nn'} \delta_{jj'} =
\end{equation*}
\begin{equation}\label{ssti.20}
= \sum_{n,j} \left| A_{n,j} \right|^2 \exp(-iE_n t/\hbar) .
\end{equation}
The Fourier transformation of Eq.~(\ref{ssti.20}) leads to the expression
\begin{equation*}
P_E = \int_{-\infty}^\infty P(t) \exp(iEt/\hbar)\, dt
\end{equation*}
\begin{equation}\label{ssti.21}
= 2\pi\hbar \sum_{n,j} \left| A_{n,j} \right|^2 \delta (E - E_n) .
\end{equation}
We see the Fourier transform of the correlation function looks like a series of $\delta$-form peaks, whose positions indicate the energy eigenvalues. The algorithm based on these ideas has been developed and successfully used for transverse energy levels search with high precision
in \cite{Shulga.13}--\cite{Shulga.b}.

Of course, the numerical integration of Eq.~(\ref{ssti.17}) could be performed for the finite time interval $T$ only. This leads to the finite width of the peaks pointing to the energy eigenvalues, and limits the method resolution. In our computations we have achieved the resolution of $E_\perp$ eigenvalues not worse than 0.001 eV.

The stationary wave functions could be obtained with the spectral method too. The superposition (\ref{ssti.18}) multiplied by $\exp \left( i E_n t / \hbar \right)$ and integrated over the time interval $T$ provides the eigenfunctions $\psi_{n, j} (x, y)$ of the eigenvalue $E_n$:
\begin{equation*}
    \int\limits_0^T \Psi (x, y, t) \exp \left( \frac{i}{\hbar} E_n t \right) dt =
\end{equation*}
\begin{equation*}
    = \sum\limits_{n', j} A_{n', j} \, \psi_{n', j} (x, y) \int\limits_0^T \exp \left[ \frac{i}{\hbar} (E_n - E_n') t \right] dt \propto
\end{equation*}
\begin{equation}\label{ssti.22}
\propto \sum\limits_{j} A_{n, j} \, \psi_{n, j} (x, y)
\end{equation}
for the large enough $T$ with a good accuracy.

The eigenfunctions computed according to (\ref{ssti.22}) by our algorithm are complex in general. To study the nodal structure it is enough to plot the positive and negative domains of the real (or imagine) part of the computed function $\psi(x,y)$, because every of the functions $\mathrm{Re}\, \psi(x,y)$ and $\mathrm{Im}\, \psi(x,y)$ are the solutions of the stationary Schr\"odinger equation due to its linearity. However, to obtain Figs. \ref{Fig5}--\ref{Fig8} we convert the eigenfunction to the pure real form
by fitting the proper global phase factor.

Initially the spectral method algorithm was implemented on the MATLAB for
easy fast prototyping and debugging. The same algorithm implementation with
minor changes allows us to calculate for the considered system both
eigenenergies spectrum for some $E_{\parallel}$ and (after that in
the separate run) eigenfunction for some already known energy level $E_n$.
Of course, direct Fourier transform appears very slow, and
unfortunately the MATLAB fast Fourier transform (FFT) performance and
memory constraints appear unsatisfactory too. So
after the code quality was proven the algorithm was reimplemented
on C language using FFTW and GSL (for complex
values calculations) libraries. The FFTW uses modern sets of floating point
instructions (SSE2, AVX) if available to work faster. The production
calculations were performed under UNIX--like operating systems on computers of
\verb|amd64| multi-processor architecture, which allow us to execute in
parallel the separate calculation runs (for different eigenfunction
calculation) each on dedicated
processor core. Note that the alternative approach of multi-thread FFTW
shows lower performance, because consumes more than one processor core only
during approximately 1/3 to 1/2 of execution time. Seems also the graphics
processor usage to do FFT will be unefficient due to costs of huge arrays
input/output between GPU and CPU reachable memories at each time step.

The algorithm parameters
for present data calculations at longitudinal energy
$E_\parallel = 1000$~MeV were following:\\
--- space lattice
 $480 \times 480$, \\
--- time step value $\Delta t = 9.2 \cdot 10^{-5}$~eV$^{-1}$, \\
--- number of time steps
 $N_t = 136621875$, \\
--- the range of $E_\perp$ values had been scanned with the energy step value $\Delta E = 5.0 \cdot 10^{-5}$~eV (for eigenenergies calculations).\\
For each of four eigenfunction symmetry classes in the double-well potential
as well as for the single-well one we compute and analyze the stationary
wave functions for about 6 to 9 energy levels at different well depths.

\section{Statistical properties of the energy spectra in regular and chaotic cases}\label{AppB}

The quantum chaos theory predicts \cite{Gutz, Schuster, Stockmann} that the distance $s$ between adjacent energy levels satisfies the Wigner distribution
\begin{equation}\label{ssti.Wigner}
p(s) = \frac{\pi s}{2D^2}\exp \left( -\frac{\pi s^2}{4D^2} \right)
\end{equation}
(where $D$ is the mean inter-level spacing) for chaotic systems that results in mutual repulsion of the energy levels, and the exponential (sometimes called as Poisson) distribution
\begin{equation}\label{ssti.exponential}
p(s) = \frac{1}{D} \exp \left( -\frac{s}{D} \right)
\end{equation}
for regular ones that results in a tendency to group the levels into shells.

\begin{figure}
\begin{center}
\includegraphics[scale=0.4]{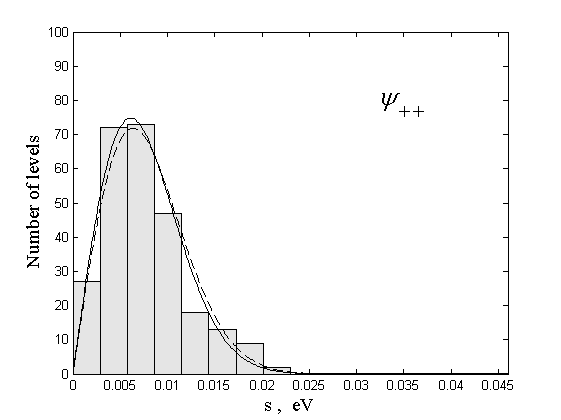} \ \includegraphics[scale=0.4]{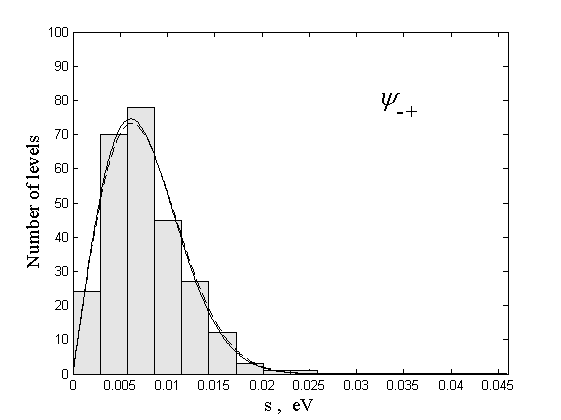} \\
\includegraphics[scale=0.4]{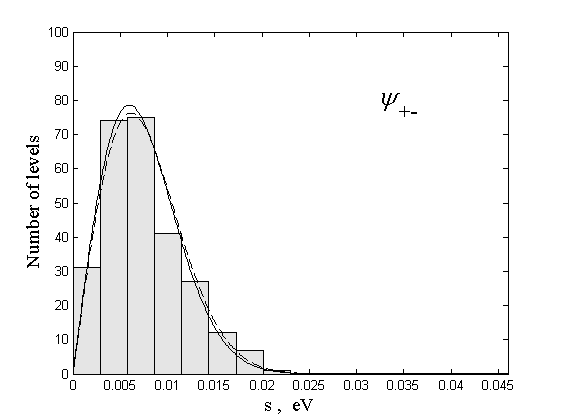} \ \includegraphics[scale=0.4]{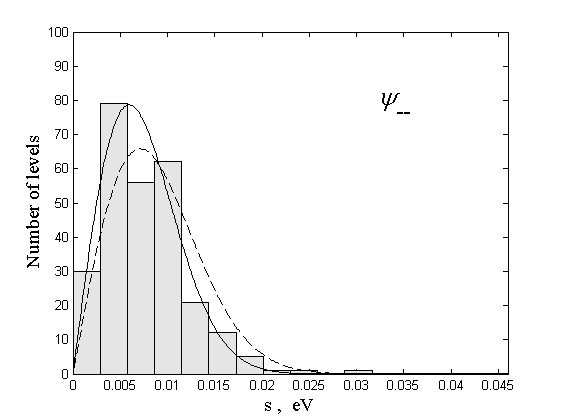} \\
\end{center}
\caption{Level spacing distributions on the $-5\leq E_\perp \leq -3$~eV interval (histograms) for (\ref{ssti.13})--(\ref{ssti.16}) types of the eigenfunction symmetry in the double-well potential. The curves show the theoretically predicted Wigner distribution (\ref{ssti.Wigner}) with the $D$ value derived from the actual data (solid lines) and obtained as a free parameter in the maximum likelihood data fit (dashed lines).}\label{Fig9}
\end{figure}

\begin{figure}
\begin{center}
\includegraphics[scale=0.4]{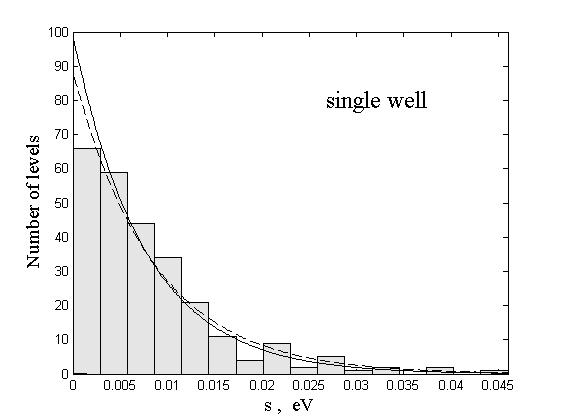} \\
\end{center}
\caption{Level spacing distribution on the $-5\leq E_\perp \leq -3$~eV interval (histogram) in the single-well potential. The curves show the theoretically predicted Poisson distribution (\ref{ssti.exponential}) with the $D$ value derived from the actual data (solid line) and obtained as a free parameter in the maximum likelihood data fit (dashed line).}\label{Fig10}
\end{figure}

The level spacing distributions in the transverse energy range $-5 \leq E_\perp \leq -3$ eV for $E_\parallel = 1$ GeV electrons channeling in the double-well potential are presented in Fig. \ref{Fig9}. Note that the statistical properties of the transverse motion eigenenergies should be investigated separately for each of four symmetry classes (\ref{ssti.13})--(\ref{ssti.16}) \cite{Stockmann}. Fig. \ref{Fig10} displays the same distribution for the energy levels in the stand-alone string potential.

The agreement between the real distributions and the theoretical predictions (\ref{ssti.Wigner}) and (\ref{ssti.exponential}), respectively, had been tested using $\chi^2$ method. The values related to this testing are given in Table \ref{ssti.Table1}. Here $N$ is the number of levels in the range under consideration, $D$ is the mean level spacing derived from the actual data, $D_{\mbox{\footnotesize fit}}$ is one obtained as a free parameter in the maximum likelihood data fit.

\begin{table}[h]
\caption{Comparison of average inter-level distance and $\chi$-square values derived from calculated data ($D$, $\chi^2$) and
obtained by fit ($D_{\mbox{\footnotesize fit}}$, $\chi^2_{\mbox{\footnotesize fit}}$).}\label{ssti.Table1}
\centering
\begin{tabular}{l c c c c c}
\hline
Symmetry type & $N$ & $D$, eV & $\chi^2$ & $D_{\mbox{\footnotesize fit}}$, eV & $\chi^2_{\mbox{\footnotesize fit}}$ \\
\hline
$\psi_{++}$ & 259 & 0.0077 & 20.6477 & 0.0080 & 17.9846 \\
$\psi_{-+}$ & 262 & 0.0076 & 23.3104 & 0.0079 & 20.8673 \\
$\psi_{+-}$ & 269 & 0.0075 & 18.5405 & 0.0077 & 16.8329 \\
$\psi_{--}$ & 261 & 0.0077 & 203.388 & 0.0088 & 41.7281 \\
single well & 235 & 0.0085 & 14.9499 & 0.0127 & 14.4778 \\
\hline
\end{tabular}
\end{table}

\end{document}